\documentstyle[prl,twocolumn,epsf,aps]{revtex} 
\tighten 
\begin{document} 
\draft 
 
\title{Universality of the 1/3 shot-noise suppression  
factor in nondegenerate diffusive conductors} 
\author{T. Gonz\'alez, C. Gonz\'alez, J. Mateos, and D. Pardo} 
\address{ 
Departamento de F\'{\i}sica Aplicada, Universidad de Salamanca, 
Plaza de la Merced s/n, \\ E-37008 Salamanca, Spain} 
\author{L. Reggiani} 
\address{ 
Istituto Nazionale di Fisica della Materia, Dipartimento di Scienza 
dei Materiali, \\ Universit\`a di Lecce, Via Arnesano, 73100 Lecce, Italy} 
\author{O. M. Bulashenko and J. M. Rub\'{\i}} 
\address{ 
Departament de F\'{\i}sica Fonamental, Universitat de Barcelona, 
Av. Diagonal 647, \\ E-08028 Barcelona, Spain} 
 
\twocolumn[ 
\date{\today} 
 
\maketitle 

\widetext \vspace*{-0.5in} 
 
\begin{abstract} 
\begin{center} \parbox{14cm}{ 
Shot-noise suppression is investigated in nondegenerate diffusive 
conductors by means of an ensemble Monte Carlo simulator. 
The universal 1/3 suppression value is obtained when transport occurs under 
elastic collision regime provided the following conditions are satisfied: 
(i) The applied voltage is much larger than 
the thermal value; (ii) the length of the device is much longer 
than both the elastic mean free path and the Debye length. 
By fully suppressing carrier-number fluctuations, 
long range Coulomb interaction  
is essential to obtain the 1/3 value in the low-frequency limit. 
} \end{center} 
\end{abstract} 
 
 
\vspace*{-1.0truecm} 
]\narrowtext 
 
In recent years kinetic phenomena in mesoscopic structures are  
offering a fascinating scenario for fundamental research 
\cite{landauer96}. 
One of the most up-to-date subjects is shot-noise suppression  
in disordered conductors. 
Here, the excess noise power has been predicted to comprise exactly  
{\it one-third} of the full shot-noise value $S_I=2eI$. 
This result has been credited to different theoretical approaches 
as applied to several microscopic models of disordered conductors. 
For a phase-coherent model Beenakker and B\"uttiker \cite{beenakker92} 
obtained the result  
using a bimodal distribution of transmission eigenvalues with the help of  
random matrix theory to calculate averages. 
For a semiclassical 1D model which includes Pauli  
principle Nagaev \cite{nagaev92}  
found the same result using a Boltzmann kinetic approach
within an elastic and energy independent relaxation-time approximation. 
For a semiclassical sequential tunneling model de Jong and Beenakker  
\cite{dejong95} 
obtained the 1/3 value within a Boltzmann-Langevin approach
in the limit of an infinite number of equal barriers and 
independently from the value of their transmission coefficient. 
Compatible results have been found by Liu {\it et al.} \cite{liu97} from a 
semiclassical implementation of a Monte Carlo simulation which includes   
Pauli principle. 
For a phase-coherent model Nazarov \cite{nazarov94} 
has proven the universality of this  
result in the diffusive limit
for arbitrary shape and resistivity  
distribution of the conductor as long as its length is  
greater than the carrier mean free path. 
Experimental evidence of the reduced shot-noise level  
close to the predicted 1/3 value in diffusive  
mesoscopic conductors has been provided in   
\cite{liefrink94,steinbach96,schoel97}. 
\par 
{}From the above it is argued that the 1/3 value of the 
suppression factor $\gamma=S_I/2eI$ is a universal phenomenon 
whose physical meaning should lay beyond  
classical or quantum mechanics and originate from some unifying concept. 
The aim of this letter is to address this issue. 
We conjecture that discreteness of charge transport is at the basis of 
such a concept, and that a transport dominated by elastic interactions 
is ultimately the physical reason for the 1/3 suppression  
independently from the quantum or classical approach used. 
Both the (apparently unrelated) coherent \cite{beenakker92} and
semiclassical \cite{nagaev92} contexts where the reduction factor 1/3
has appeared assume a degenerate Fermi gas, and the noise reduction comes
from the regulation of electron motion by the exclusion principle.
Landauer claims that the appearance of 1/3 in these very different cases 
is a numerical coincidence \cite{landauer96}.
We argue that {\em neither phase coherence nor Fermi statistics are required 
for the occurrence of suppressed shot noise in diffusive conductors}.
We show a third case where the origin of the effect is completely 
classical, and the correlation between electrons comes from their
Coulomb interaction, rather than the exclusion principle. The repeated
occurrence of 1/3 leads us to believe that this must be more 
than a numerical coincidence.
\par 
To support the above conjecture, we present 
the results of a Monte Carlo simulation for a nondegenerate  
diffusive conductor. 
By providing an exact solution of the kinetic equation coupled with 
a Poisson solver (PS) we are free from any approximation and thus 
in the position to obtain a rigorous proof of our findings. 
Furthermore, the role of long-range Coulomb interaction 
\cite{shot1,shot3}, although known since the times of vacuum diodes
\cite{vacuum}, is here considered for the relevant case of a medium
in the presence of elastic and inelastic scattering for the
first time. 
\par
For the calculations we consider the following simple model:  
a lightly doped active region of a semiconductor sample of length $L$ 
sandwiched between two heavily doped contacts 
injecting carriers into the active region. 
The contacts are considered to be Ohmic (the voltage drop inside them being 
negligible) and to remain always at thermal equilibrium. 
Thus, electrons are emitted from the contacts according to  
a thermal-equilibrium Maxwell-Boltzmann distribution at the lattice  
temperature $T$, and they move inside the active region  
following the classical equations of motion by undergoing 
isotropic scattering in momentum space. 
To exclude additional correlations due to Fermi statistics,  
the electron gas is assumed to be nondegenerate. 
In principle, electrons are injected with a Poissonian statistics, i.e., 
the time between two consecutive electron emissions 
is generated with a probability $P(t)=\Gamma e^{-\Gamma t}$, 
where $\Gamma=\frac{1}{2}n_c v_{\rm th}S$ is the injection-rate density, 
$v_{\rm th}=\sqrt{2k_B T/\pi m}$ the thermal velocity,  
$S$ the cross-sectional 
area of the device, and $m$ the electron effective mass. 
However, to prove our conjecture that the origin of $\gamma=1/3$ is 
just elastic scattering and therefore independent of the contact 
injection, the fluctuating emission rate at the contacts  
in the diffusive limit is taken to follow other kind of statistics 
ranging from uniform to Poissonian. 
For the simulations we have used the following set of parameters: 
$T=300\,K$, $m=0.25\,m_0$ ($m_0$ being the free electron mass), 
relative dielectric constant $\varepsilon$=11.7, $L=200 \ nm$, and  
$n_c=4\times 10^{17} {\rm cm}^{-3}$ (much higher than the sample doping). 
The above set of values yields for the dimensionless parameter 
$\lambda=L/L_{Dc}$ (with $L_{Dc}$ the Debye length corresponding 
to $n_c$), which characterizes the importance 
of the electrostatic screening \cite{shot1}, the value $\lambda=30.9$, 
which implies significant space-charge effects and inhomogeneity  
inside the structure. 
Furthermore, the average time between collisions in the bulk $\tau$  
is assumed to be independent of energy, and  
is varied from $10^{-15} {\rm s}$ to $10^{-11} {\rm s}$, so that both regimes 
of carrier transport, ballistic ($\ell /L \gg 1$) and diffusive 
($\ell /L \ll 1$), are covered. In our calculations $\ell$, the  
carrier mean free path, is estimated as $v_{\rm th}\tau$. 
To analyze the effect of scattering inelasticity, collisions are treated  
as elastic or as inelastic. In the latter case the carrier is thermalized 
after each scattering event. 
\par 
We apply a $dc$ voltage and calculate the time-averaged 
current $I$ and the current autocorrelation function $C_I(t)$ 
by means of an ensemble Monte Carlo simulator self-consistently 
coupled with a PS.  
We assume that the active region of the structure in transversal directions  
is sufficiently thick to allow a 1D electrostatic treatment. 
Accordingly, the simulation is 1D in real space  
and 3D in momentum space. 
We stress that in our approach the number of electrons $N$ inside  
the sample 
fluctuates in time due to the random injection/extraction from  
the contacts, and we can evaluate both the time-averaged value  
$\langle N \rangle$ and its fluctuations. 
To analyze the importance of the effects associated with  
long-range Coulomb interaction, 
we provide the results for two different simulation schemes \cite{shot1}. 
The first one involves a {\em dynamic} PS,
where the potential is self-consistently updated  
at each time step during the simulation 
by solving the Poisson equation
under the condition that the contact potential remains time independent. 
The second scheme makes use of a {\em static} PS, so that only  
the stationary potential profile is calculated  
and carriers move in the {\em frozen} non-fluctuating electric field profile. 
{\em Both schemes give exactly the same average current and steady-state  
spatial distributions of all the quantities, but the noise  
characteristics are in general quite different.} 
\par 
Figure \ref{baldif} presents the results for the low-frequency 
suppression factor $\gamma = S_I/2eI$ as a function of $\ell /L$ 
for an applied voltage of $U=40k_B T/e$. 
When the transport is ballistic or quasi-ballistic 
($\ell/L \gtrsim 10^{-1}$), the static results 
show full shot noise ($\gamma \approx 1$ within numerical uncertainty)
associated with the carrier injection (which in this range
is modeled as Poissonian),  
whereas in the case of the dynamic results  
space charge is responsible for a relevant noise suppression, 
yielding for the suppression factor the value of 0.045 
under perfect ballistic regime ($\ell / L \gg 1$) \cite{shot1}.  
The observed reduced shot-noise level is accompanied by a 
sub-Poissonian electron number statistics \cite{shot3}.
In the transition from ballistic to diffusive regime,  
the suppression due to long-range Coulomb interaction 
remains active, being more pronounced in the case 
of inelastic scattering mechanisms. 
Under fully diffusive regime, attained at  
$\ell/L \lesssim 10^{-2}$, the results become independent of the
carrier injecting statistics and
the following asymptotic conditions are detected. 
The static results, both in the elastic and inelastic cases, keep
the full shot-noise limit. 
The elastic-dynamic case attains the 1/3 value 
within numerical uncertainty in excellent agreement with theoretical  
expectations. 
The inelastic-dynamic case is further suppressed well below 1/3. 
\par 
Figure \ref{baldif2} complements the results of the elastic dynamic case in 
Fig.\ \ref{baldif} by presenting calculations at increasing applied voltages. 
Here we find the remarkable fact that, for high voltages,
the 1/3 value in the diffusive limit is reached
from the full shot-noise value in the ballistic limit. 
The reason for the different behavior in the ballistic regime
is the presence or absence of a potential barrier near
the cathode which controls the current through the structure. 
For the highest voltages, when the barrier disappears, the current saturates  
and the suppression factor takes on the full shot-noise level  
as for the static case \cite{shot1}. 
It has been checked that the onset of the 1/3 value takes place when,
because of intensive elastic scattering,
a significant energy redistribution among the three velocity components
is achieved.
Accordingly, the higher the applied voltage, the larger the  
energy gained by electrons in the free flights and, consequently, 
more scattering is needed to redistribute the energy  
equally among the components.  
Hence, the threshold of the 1/3 regime is shifted 
towards lower values of $\ell/L$ as the applied voltage increases. 
We notice that the 1/3 limit exhibits several  
{\em universal} properties, namely, it is independent of: 
(i) the applied voltage bias (see Fig.\ \ref{baldif2}), 
(ii) the scattering strength, (iii) and the carrier injecting statistics. 
\par
To illustrate and explain the physical origin of the 1/3 value,
Fig.\ \ref{spec} reports a typical frequency spectrum of the suppression
factor $S_I(f)/2eI$ under elastic-diffusive conditions 
for static and dynamic PS. Monte Carlo simulation
allows the calculation of the three terms in which $S_I(f)$ can be
decomposed, namely velocity, number, and cross-correlation contributions  
$S_I(f)=S_V(f)+S_N(f)+S_{VN}(f)$ \cite{shot1}.  
In the static case the spectrum clearly shows that all three terms
contribute to $S_I(f)$, and two different time scales can be identified.
The longest one, associated with the transit time of the carriers
through the device $\tau _T$, is evidenced in the terms $S_N$ and
$S_{VN}$. The shortest one, related to the relaxation time of
elastic scattering $\tau$, is manifested in $S_V$. The latter is responsible
for 1/3 (within the numerical uncertainty) of the full shot-noise value
obtained for the suppression factor at low frequency, while the former 
provides the rest of the contribution up to the value 1. 
In contrast with this behavior, in the dynamic case
the number contribution is found to be compensated by a negative
velocity-number cross-correlation contribution and,
as a result, $S_I(f)$ is found to coincide with $S_V(f)$ 
in all the frequency range, thus showing only the cut-off at
higher frequencies. 
It is important to notice that in the frequency range
${1\over{2\pi\tau_T}}\lesssim f \lesssim {1\over{2\pi\tau}}$ 
both static and dynamic suppression factors exhibit the 1/3 value, 
which is related to velocity fluctuations. 
However, at low frequency only the dynamic case takes this value 
by virtue of Coulomb correlations, which are responsible for
the mutual compensation of $S_N$ and $S_{VN}$ contributions.
For this compensation to take place we have checked that it is necessary 
to fulfil the condition $L \gg L_{Dc}$
in order to achieve a significant action of long-range Coulomb interaction.
\par 
Figure\ \ref{dif} reports the low-frequency suppression factor 
in the diffusive regime ($\ell / L = 10^{-3}$) 
as a function of the applied voltage, calculated by using the 
dynamic PS. We stress that these results are independent of
the injecting statistics at the contacts.
By comparing the elastic and inelastic cases we find that at the 
lowest bias both cases coincide by providing the standard thermal noise as 
predicted by Nyquist relation. 
On the contrary, at the highest voltages the elastic case achieves the 
1/3 limiting value, in full agreement with our conjecture, while the  
inelastic case continues decreasing. 
\par 
The results for the low-frequency value of the spectral density  
in the inelastic case $S_I^{inel}$ are closely fitted  
by the simple relation: 
\begin{equation}\label{siinel} 
S_I^{inel}=4k_BTG_0{\langle N \rangle\over\langle N \rangle_0} 
\end{equation} 
where $G_0$ is the conductance and $\langle N \rangle_0$ the average number 
of electrons inside the sample, both in the limit of vanishing bias. 
This result means that, under strong inelastic scattering, the noise in our 
sample is just the thermal Nyquist noise (modulated by the  
variation in the number of  
carriers) even in the presence of a high bias and a net current 
flowing through the sample.  
Therefore, inelastic scattering 
strongly suppresses shot noise and makes the noise become 
macroscopic ($\gamma \ll 1$, see Fig.\ \ref{dif}). 
These results confirm previous predictions 
by Liu and Yamamoto \cite{Liu94} and Nagaev \cite{Nagaev95}, with the 
important difference that our approach is completely classical. 
The conclusion of Shimizu and Ueda \cite{shimizu92} that, while dephasing
is irrelevant,
energy transfer from electrons to other systems is essential for shot-noise
suppression 
is also supported by present results. 
\par
The values of the low-frequency spectral density in the elastic case 
$S_I^{el}$ are nicely reproduced by the expression:
\begin{equation}\label{siel}
S_I^{el}={8\over3}k_BTG_0{\langle N \rangle\over\langle N \rangle_0}
+{2\over3}eI\coth \left({eU \over 2k_BT}\right)
\end{equation}
which is quite similar to that obtained by Nagaev \cite{nagaev92}, 
and describes the crossover
from thermal-Nyquist noise for $eU\ll k_BT$ ($S_I^{el}=4k_BTG_0$) to
suppressed shot noise for $eU\gg k_BT$ ($S_I^{el}={2\over3}eI$), thus
providing $\gamma=1/3$ for the highest applied voltages (Fig.\ \ref{dif}).
The main difference between the effect of elastic and inelastic 
scattering on the noise in our model is the following. 
Strong inelastic scattering, by dissipating the energy that electrons gain
from the field and randomizing the momentum,
reduces the electron average energy to that of the lattice, and
therefore the noise corresponds to thermal-Nyquist noise at any bias.
On the other hand, elastic scattering, by simply randomizing the
electron momentum, allows for the broadening of the electron
velocity distribution at high voltages.
As a consequence, the level of noise increases with respect to
equilibrium conditions, which results in
a noise power with value $1/3$ of the full shot-noise value \cite{remark1}.
In addition, we have checked that if a 2D momentum space is
considered (electron energy is shared between two velocity
components after each scattering), the observed 
$\gamma=S_I^{el}/2qI$ in diffusive regime is about {\em one-half},
which indicates that the suppression factor is related to
the momentum-space dimensionality \cite{remark2}.
\par
In conclusion, within an ensemble Monte Carlo scheme  
we have investigated the shot-noise suppression in nondegenerate diffusive 
conductors. 
Results prove that the 1/3 value of the suppression factor  
is related to a transport dominated by elastic scattering (diffusive limit) 
under the condition $eU \gg k_BT$. 
The appearance of this factor requires the simultaneous validity of  
the two conditions $\ell \ll L$ and $L_{Dc} \ll L$, 
the former implying the achievement of fully diffusive conditions, the latter 
the decisive role of long range Coulomb correlations in suppressing  
the contribution associated with number fluctuations. 
Inelastic scattering is found to strongly suppress shot noise,  
reducing it to thermal Nyquist noise under heavily dissipative conditions. 
\par 
This work has been partially supported by the Comi\-si\'on Interministerial de 
Ciencia y Tecnolog\'{\i}a through Project TIC95-0652 and Ministero  
dell' Universit\`a e della Ricerca Scientifica e Tecnologica (MURST). 
O.M.B. acknowledges support by the Direcci\'on General de Ense\~nanza  
Superior, Spain.

\begin{figure} 
\setlength{\epsfxsize}{8cm} 
\centerline{\mbox{\epsffile{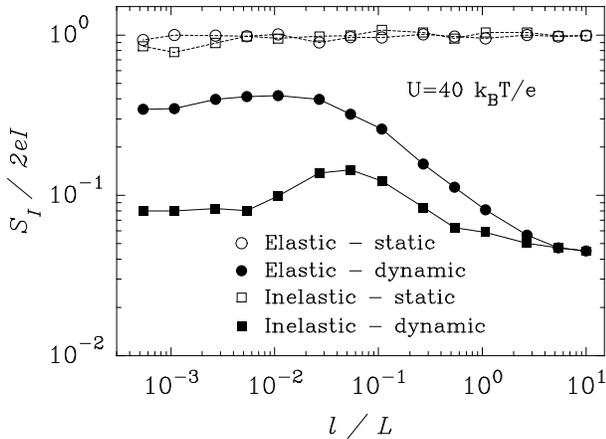}}}  
\caption{ 
Shot-noise suppression factor vs ballistic parameter  
$\ell /L $ for an applied voltage of $U=40k_B T/e$ 
calculated by using static (open symbols, dashed line) and dynamic  
(full symbols, solid line) potentials and elastic (circles)  
and inelastic (squares) scattering mechanisms. 
}\label{baldif}\end{figure} 
 
\begin{figure} 
\setlength{\epsfxsize}{8cm} 
\centerline{\mbox{\epsffile{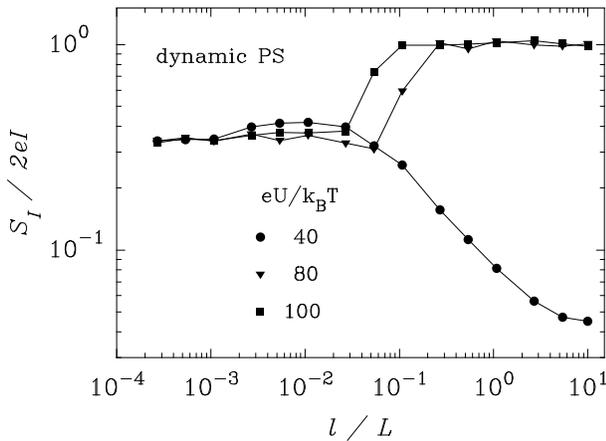}}} 
\caption{ 
Shot-noise suppression factor vs ballistic parameter  
$\ell /L $ for the case of elastic scattering mechanisms 
at different applied voltages. 
Calculations are performed by using dynamic (self-consistent) potential. 
}\label{baldif2}\end{figure} 
 
\begin{figure} 
\setlength{\epsfxsize}{8cm} 
\centerline{\mbox{\epsffile{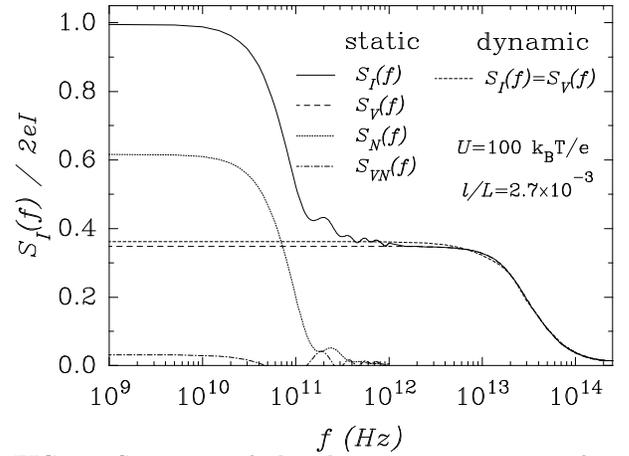}}} 
\caption{ 
Spectrum of the shot-noise suppression factor for the case of static and  
dynamic potentials under elastic-diffusive conditions. 
Different contributions to the total value are indicated in the figure. 
}\label{spec}\end{figure} 
 
\begin{figure} 
\setlength{\epsfxsize}{8cm} 
\centerline{\mbox{\epsffile{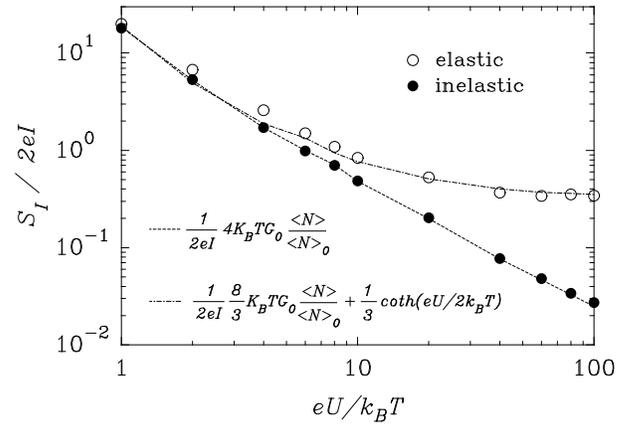}}}  
\caption{ 
Shot-noise suppression factor vs applied bias $U$  
calculated with dynamic potential for the cases of elastic and  
inelastic scattering  
mechanisms with $\ell / L = 10^{-3}$ and 
an injecting concentration $n_c = 4\times 10^{17} \rm cm^{-3}$. 
The curves correspond to the fittings of Eqs.\ (1) and (2). 
}\label{dif}\end{figure} 
 

\begin{references}
\bibitem{remark2}
The appearance of the 1/3 factor in the 1D simulation of Ref.\cite{liu97},
which is in apparent contradiction with this conclusion, is due to
the effect of Pauli exclusion principle, while in our case the correlation
between electrons comes from their Coulomb interaction.
\end{references}
\end{document}